\documentstyle[twocolumn,aps,epsfig]{revtex}

\begin{document}
\twocolumn[\hsize\textwidth\columnwidth\hsize\csname@twocolumnfalse\endcsname

\draft
\title{Formation of energy gap in higher dimensional spin-orbital liquids}
\author{Shun-Qing Shen$^{1}$ and Guang-Ming Zhang$^{2}$}
\address{$^{1}$Department of Physics, the University of Hong Kong, Hong Kong, China\\
$^{2}$Center for Advanced Study, Tsinghua University, Beijing 100084, China}
\date{\today}
\maketitle

\begin{abstract}
A Schwinger boson mean field theory is developed for spin liquids in a
symmetric spin-orbital model in higher dimensions. Spin, orbital and coupled
spin-orbital operators are treated equally. We evaluate the dynamic
correlation functions and collective excitations spectra. As the collective
excitations have a finite energy gap, we conclude that the ground state is a
spin-orbital liquid with a two-fold degeneracy, which breaks the discrete
spin-orbital symmetry. Possible relevence of this spin liquid state to
several realistic systems, such as CaV$_4$V$_9$ and Na$_2$Sb$_2$Ti$_2$O, are
discussed.
\end{abstract}


]

The formation of a spin gap in two- or higher dimensional quantum spin
systems is a long-standing issue in strongly correlated problems. \cite
{Dagotto96} Several physical mechanisms were proposed to explain the
formation of the spin gap in the low-energy excitations. Most of them focus
on one-dimensional spin chains and spin ladders, such as an S=1
antiferromagnetic Heisenberg chain and even-leg spin ladders. In higher
dimensions Anderson proposed that strong quantum fluctuations for spin-1/2
systems may destroy the antiferromagnetic long-range order in two dimension,
and lead to form a resonating valence bond (RVB) state where a spin gap may
open. \cite{Anderson73} However, it becomes true only for some frustrated
spin systems such as on the Kagome lattice or in the Majumdar-Ghosh model
with a strong next nearest neighbor interaction, otherwise there exists
antiferromagnetic long-range orders in the ground state on a square and
triangle lattice. Recently it has been realized that orbital degrees of
freedom of d- and f-electrons in transition metal ions provide a new route
to find the physical mechanism of spin gap formation. Several spin-orbital
models \cite{Santoro97,Pati98,Li98,Martins00,Ito00,Bossche01} have shown the
tendency of the formation of a spin gap in the ground states due to strong
orbital and spin quantum fluctuations. Accumulating numerical calculations
show that spin liquid state may be formed in some one-dimensional
spin-orbital coupled systems. Behaviors in higher dimensional systems are
relatively less clear. There are several higher dimensional spin-gap
materials such as Na$_2$Ti$_2$Sb$_2$O \cite{Axtell97} and NaV$_4$O$_9$\cite
{Taniguchi95}, in which the orbital degree of freedom might play a key role
in the formation of the spin gap.

The spin-orbital model Hamiltonian is written as \cite{Kugel73,Castellani78}

\begin{eqnarray}
H &=&J\sum_{\left\langle ij\right\rangle }\left( 2{\bf S}_{i}\cdot {\bf S}%
_{j}+\frac{1}{2}\right) \left( 2\tau _{i}\cdot \tau _{j}+\frac{1}{2}\right) 
\nonumber \\
&&+J_{s}\sum_{\left\langle ij\right\rangle }\left( 2{\bf S}_{i}\cdot {\bf S}%
_{j}+\frac{1}{2}\right) +J_{\tau }\sum_{\left\langle ij\right\rangle }\left(
2\tau _{i}\cdot \tau _{j}+\frac{1}{2}\right) .
\end{eqnarray}
This model may be derived from an electronic model with double orbital
degeneracy. This model possesses an $SU(2)\otimes SU(2)$ symmetry. When $%
J_{s}=J_{\tau }=J_{0}$, the model have an additional discrete symmetry,
namely, the permutation symmetry between spin and orbital. In this case the
model can be written as a combination of two symmetric models 
\begin{eqnarray}
H &=&(J+J_{0})\sum_{\left\langle ij\right\rangle }\left( 2{\bf S}_{i}\cdot 
{\bf S}_{j}+\frac{1}{2}\right) \left( 2\tau _{i}\cdot \tau _{j}+\frac{1}{2}%
\right)  \nonumber \\
&&-J_{0}\sum_{\left\langle ij\right\rangle }\left( 2{\bf S}_{i}\cdot {\bf S}%
_{j}-\frac{1}{2}\right) \left( 2\tau _{i}\cdot \tau _{j}-\frac{1}{2}\right) .
\end{eqnarray}
The former part is the standard SU(4) spin-orbital model, which is solvable
in one-dimension and has been investigated intensively. \cite{Li98} The
second part is the model first proposed by Santoro\cite{Santoro97}, and also
possesses the SU(4) symmetry with different generators. However the
combination of the two models breaks the SU(4) symmetry.

>From previous studies, it has been found that the interplay between spin
and orbital degrees of freedom produces either quantum ordered or disordered
phases. Spin liquid states with an energy gap were found in one dimensions. 
\cite{Santoro97,Pati98,Ito00} To investigate the model systematically, we
try to develop a simple theory, which can describe the disordered state with
an energy gap as well as the ordered states. The Schwinger boson theory is
an ideal candidate. The theory was first used to the spin SU(2) Heisenberg
model, and then was generalized to SU(N) systems.\cite{Auerbach88} The
advantage of this theory is that it can describe either quantum ordered or
disordered states. The results for two- and three-dimensional Heisenberg
model are consistent with the spin wave theory very well. Here we present
the Schwinger boson mean field theory for this spin-orbital system.

For the present model, there are four possible states on each site $i$
according to the eigenvalues of ${\bf S}_i^z$ and $\tau _i^z$: $\left|
1\right\rangle =\left| +1/2,+1/2\right\rangle $, $\left| 2\right\rangle
=\left| -1/2,+1/2\right\rangle ,$ $\left| 3\right\rangle =\left|
+1/2,-1/2\right\rangle ,$ $\left| 4\right\rangle =\left|
-1/2,-1/2\right\rangle .$ We introduce four Schwinger bosons to describe
these four states: $\left| \mu \right\rangle =a_{i\mu }^{\dagger }\left|
0\right\rangle ,$ where $\left| 0\right\rangle $ is the vacuum states and $%
\mu =1,2,3,4$. There is a constraint for the four bosons, $\sum_{\mu
=1}^4a_{i\mu }^{\dagger }a_{i\mu }=1$, for each site. On these basis the
spin and orbital operators can be expressed in terms of these four Schwinger
bosons. 
\begin{eqnarray*}
S_i^{+} &=&a_{i1}^{\dagger }a_{i2}+a_{i3}^{\dagger }a_{i4};\text{ \ }\tau
_i^{+}=a_{i1}^{\dagger }a_{i3}+a_{i2}^{\dagger }a_{i4}; \\
S_i^{-} &=&a_{i2}^{\dagger }a_{i1}+a_{i4}^{\dagger }a_{i3};\text{ \ }\tau
_i^{-}=a_{i3}^{\dagger }a_{i1}+a_{i4}^{\dagger }a_{i2}; \\
S_i^z &=&\frac 12\left( a_{i1}^{\dagger }a_{i1}-a_{i2}^{\dagger
}a_{i2}+a_{i3}^{\dagger }a_{i3}-a_{i4}^{\dagger }a_{i4}\right) ; \\
\tau _i^z &=&\frac 12\left( a_{i1}^{\dagger }a_{i1}-a_{i3}^{\dagger
}a_{i3}+a_{i2}^{\dagger }a_{i2}-a_{i4}^{\dagger }a_{i4}\right) .
\end{eqnarray*}
Thus, the Hamiltonian is rewritten in terms of the Schwinger bosons 
\begin{eqnarray}
H &=&-\frac{J+J_0}2\sum_{\left\langle ij\right\rangle ,\mu \nu }A_{ij,\mu
\nu }^{\dagger }A_{ij,\mu \nu }  \nonumber \\
&&-J_0\sum_{\left\langle ij\right\rangle ,\mu \nu }(B_{ij,14}^{\dagger
}-B_{ij,23}^{\dagger })(B_{ij,14}-B_{ij,23})  \nonumber \\
&&+\sum_i\lambda _i(\sum_{\mu =1}^4a_{i\mu }^{\dagger }a_{i\mu }-1)+\frac 12%
zN_\Lambda (J+2J_0)
\end{eqnarray}
where $A_{ij,\mu \nu }=a_{i\mu }a_{j\nu }-a_{i\nu }a_{j\mu }$ and $B_{ij,\mu
\nu }=a_{i\mu }a_{j\nu }+a_{i\nu }a_{j\mu }.$ We have introduced
antisymmetric and symmetric operators A and B for the purpose of the mean
field calculations. The following theory is limited to the case $-J\geq
J_0\geq 0.$ We should introduce different order parameters in the different
parameter range. The local Lagrangian multiplier is introduced to realize
the local constraint for hard core bosons. In the mean field approach we
shall take it as site-independent $\lambda $. The thermodynamic averages of
the operators A and B are introduced as the order parameters, respectively, 
\[
\left\langle A_{ij,\mu \nu }\right\rangle \equiv -2i\Delta _{\mu \nu
}^o(r_i-r_j);\ \left\langle B_{ij,\mu \nu }\right\rangle \equiv 2\Delta
_{\mu \nu }^e(r_i-r_j). 
\]
$\Delta _{\mu \nu }^o(r_i-r_j)$ and $\Delta _{\mu \nu }^e(r_i-r_j)$ are odd
and even functions with respect to the indices $\mu $, $\nu $ or the sites $%
r_i$, $r_j.$ In the momentum space, we take 
\begin{eqnarray*}
\frac iZ\sum_\delta \Delta _{\mu \nu }^o(\delta )e^{-ik\cdot \delta }
&\equiv &\frac{\ \Delta _{\mu \nu }^o}d\sum_\alpha \sin k_\alpha \equiv \
\Delta _{\mu \nu }^o\gamma _s(k); \\
\frac 1Z\sum_\delta \Delta _{\mu \nu }^e(\delta )e^{-ik\cdot \delta }
&\equiv &\frac{\ \Delta _{\mu \nu }^e}d\sum_\alpha \cos k_\alpha \equiv \
\Delta _{\mu \nu }^e\gamma _c(k)
\end{eqnarray*}
where $\delta $ points to the nearest neighbor sites. By utilizing the Pauli
matrices $\sigma _\alpha $ ($\alpha =x,y,z$ ) and the identity matrix $%
\sigma _0$, the decoupled Hamiltonian can be expressed in a compact form of $%
8\times 8$ matrix, 
\[
H=\frac 12\sum_k\Phi _k^{\dagger }H(k)\Phi _k+{\cal E}_0 
\]
where 
\begin{eqnarray*}
\Phi _k^{\dagger } &=&(a_{k1}^{\dagger },a_{k2}^{\dagger },a_{k3}^{\dagger
},a_{k4}^{\dagger },a_{-k1},a_{-k2},a_{-k3},a_{-k4}); \\
H(k) &=&\lambda \sigma _0\otimes \sigma _0\otimes \sigma _0-i\sigma
_y\otimes A(k)+b(k)\sigma _x\otimes \sigma _y\otimes \sigma _y; \\
A(k) &=&-2z(J+J_0)\gamma _s(k)\left( 
\begin{array}{cccc}
0 & \Delta _{12}^o & \Delta _{13}^o & \Delta _{14}^o \\ 
-\Delta _{12}^o & 0 & \Delta _{23}^o & \Delta _{24}^o \\ 
-\Delta _{13}^o & -\Delta _{23}^o & 0 & \Delta _{34}^o \\ 
-\Delta _{14}^o & -\Delta _{24}^o & -\Delta _{34}^o & 0
\end{array}
\right) ; \\
\frac{{\cal E}_0}{N_\Lambda } &=&+z(J+J_0)\sum \Delta _{\mu \nu
}^2+2zJ_0(\Delta _{14}^e-\Delta _{23}^e)^2 \\
&&-3\lambda +\frac 12z(J+2J_0),
\end{eqnarray*}
where $b(k)=-2zJ_0(\Delta _{14}^e-\Delta _{23}^e)\gamma _c(k)$. The
Kronecker product for block matrices is used.\cite{Golub96} The Hamiltonian
can thus be diagonalized analytically, and the free energy is evaluated to
establish the mean field equations. Due to the symmetry in the Hamiltonian,
there exists two sets of solutions: (I) $\Delta _{14}^e-\Delta
_{23}^e=\Delta _2$, $\Delta _{12}^o=\Delta _{34}^o=\Delta ,$ $\Delta
_{13}^o=-\Delta _{24}^o,$ $\Delta _{14}^o=\Delta _{23}^o$ with $\Delta _1=%
\sqrt{\Delta _{13}^2+\Delta _{14}^2};$ (II) $\ \Delta _{14}^e-\Delta
_{23}^e=\Delta _2$, $\Delta _{12}^o=-\Delta _{34}^o,$ $\Delta _{13}^o=\Delta
_{24}^o,$ $\Delta _{14}^o=-\Delta _{23}^o=\Delta $ with $\Delta _1=\sqrt{%
\Delta _{12}^2+\Delta _{13}^2}.$ With the notation $\Delta ,$ $\Delta _1,$
and $\Delta _2,$ the same branches of spectra are given by

\begin{equation}
\omega (k)_{\pm }=\sqrt{\lambda ^2-a^2\left( k\right) -b_{\pm }^2(k)},
\end{equation}
where $a(k)=-2z(J+J_0)\Delta \gamma _s(k)$ and $b_{\pm }(k)=-2z(J+J_0)\Delta
_1\gamma _s(k)\pm 2zJ_0\Delta _2\gamma _c(k)$. Moreover, these two ground
states are degenerated. The degeneracy originates from the symmetry of
permutation of spin and orbital operators. When $J_0=0,$ the model is
reduced to the standard SU(4) spin-orbital model. The two spectra become
degenerated, 
\[
\omega (k)=\sqrt{\lambda ^2-[2z(J+J_0)\gamma _s(k)]^2(\Delta _{12}^2+\Delta
_{13}^2+\Delta _{14}^2)}, 
\]
This way we recover the spectra for the SU(N=4) model.

In the following we will focus on the ground state with an energy gap. So we
do not consider the Bose-Einstein condensation which may give rise to the
long-range order. Indeed the long-range orders appear when $J_0=0$ or $%
J=-J_0 $ on two- and three-dimensional hypercubic lattices.\cite{Zhang01}
The condensation occurs when the system deviates from those symmetric point
slightly. General description of complete solutions to the problem will be
presented elsewhere. To determine the order parameters, we introduce a set
of dimensionless parameters $\tilde{\Delta}=$ $2z(J+J_0)\Delta /\lambda ,$ $%
\tilde{\Delta}_1=$ $2z(J+J_0)\Delta _1/\lambda ,$ $\tilde{\Delta}_2=$ $%
2zJ_0\Delta _2/\lambda ,$ and $\tilde{\lambda}=\lambda /[z(J+J_0)]$, then
the quasiparticle excitation spectrum becomes 
\begin{equation}
\tilde{\omega}_{\pm }(k)=\sqrt{1-\tilde{\Delta}^2\gamma _s^2(k)-[\tilde{%
\Delta}_1\gamma _s(k)\pm \tilde{\Delta}_2\gamma _c(k)]^2}.
\end{equation}
The self-consistent mean field equations are given by 
\begin{eqnarray}
\int \frac{dk}{(2\pi )^d}\left[ \frac 1{\tilde{\omega}_{+}(k)}+\frac 1{%
\tilde{\omega}_{-}(k)}\right] \ &=&3  \label{mf1} \\
\int \frac{dk}{(2\pi )^d}\left[ \frac{\gamma _s^2(k)}{\tilde{\omega}_{+}(k)}%
\ +\frac{\gamma _s^2(k)}{\tilde{\omega}_{-}(k)}\right] &=&2\tilde{\lambda}
\label{mf2} \\
\int \frac{dk}{(2\pi )^d}\left[ 
\begin{array}{c}
\frac{(\tilde{\Delta}_1\gamma _s(k)+\tilde{\Delta}_2\gamma _c(k))\gamma _s(k)%
}{\tilde{\omega}_{+}(k)}\  \\ 
+\frac{(\tilde{\Delta}_1\gamma _s(k)-\tilde{\Delta}_2\gamma _c(k))\gamma
_s(k)}{\tilde{\omega}_{-}(k)}
\end{array}
\right] &=&2\tilde{\lambda}\tilde{\Delta}_1  \label{mf3} \\
\int \frac{dk}{(2\pi )^d}\left[ 
\begin{array}{c}
\frac{(\tilde{\Delta}_1\gamma _s(k)+\tilde{\Delta}_2\gamma _c(k))\gamma _c(k)%
}{\tilde{\omega}_{+}(k)}\  \\ 
-\frac{(\tilde{\Delta}_1\gamma _s(k)-\tilde{\Delta}_2\gamma _c(k))\gamma
_c(k)}{\tilde{\omega}_{-}(k)}
\end{array}
\right] &=&\tilde{\lambda}\tilde{\Delta}_2(1+\frac J{J_0}).  \label{mf4}
\end{eqnarray}
Substituting Eq.(\ref{mf2}) into Eq.(\ref{mf3}), we have 
\[
\int \frac{dk}{(2\pi )^d}\left[ \frac 1{\tilde{\omega}_{+}(k)}\ -\frac 1{%
\tilde{\omega}_{-}(k)}\right] \tilde{\Delta}_2\gamma _c(k)\gamma _s(k)=0. 
\]
If $\tilde{\Delta}_2\neq 0,$ $\tilde{\Delta}_1$ must be equal to zero.
Oppositely, if $\tilde{\Delta}_2=0,$ \ the solution is for the case of $%
J_0=0.$ Therefore for $J_0>0,$ the solution is $\tilde{\Delta}_1=0$ with $%
\tilde{\omega}_{+}(k)=\tilde{\omega}_{-}(k)\equiv \tilde{\omega}(k).$ The
two spectra are also degenerated. Two sets of saddle point solutions become,
corresponding to the spin liquid phase with an energy gap in elementary
excitations: (I) $\Delta _{14}^e-\Delta _{23}^e=\Delta _2$, $\Delta
_{12}^o=\Delta _{34}^o=\Delta ,$ and $\Delta _{\mu \nu }^o=0$ otherwise.
(II) $\ \Delta _{14}^e-\Delta _{23}^e=\Delta _2$, $\Delta _{13}^o=-\Delta
_{24}^o=\Delta ,$ and $\Delta _{\mu \nu }^o=0$ otherwise. We focus on the
first set of solutions and then present the results for the second set of
solutions.

The one-particle Green's function in an $8\times 8$ matrix form is defined
as 
\[
{\bf G}(k,t)=-i\left\langle 0|T\{\Phi _k(t)\Phi _k^{\dagger
}(0)\}|0\right\rangle 
\]
where $\left| 0\right\rangle $ is the ground state of the Hamiltonian. Its
Fourier transform is given by 
\begin{eqnarray*}
{\bf G}(k,\omega ) &=&\left( (\omega +i\delta )\sigma _z\otimes \sigma
_0\otimes \sigma _0-H_k\right) ^{-1} \\
&=&\frac 1{\omega ^2-\lambda ^2+a^2\left( k\right) +b^2(k)+i\delta }\times \\
&&\left\{ 
\begin{array}{c}
\omega \sigma _z\otimes \sigma _0\otimes \sigma _0-\lambda \sigma _0\otimes
\sigma _0\otimes \sigma _0 \\ 
+a(k)\sigma _y\otimes \sigma _0\otimes \sigma _y-b(k)\sigma _x\otimes \sigma
_y\otimes \sigma _y
\end{array}
\right\} .
\end{eqnarray*}
with $a(k)=-2z(J+J_0)\Delta \gamma _s(k)$ and $b(k)=-2zJ_0\Delta _2\gamma
_c(k).$ The corresponding saddle point equations then become: 
\begin{eqnarray*}
\int \frac{dk}{(2\pi )^d}\frac 1{\tilde{\omega}(k)}\ &=&\frac 32;\text{ } \\
\int \frac{dk}{(2\pi )^d}\frac{\gamma _s^2(k)}{\tilde{\omega}(k)}\ &=&\tilde{%
\lambda}; \\
\int \frac{dk}{(2\pi )^d}\frac{\gamma _c^2(k)}{\tilde{\omega}(k)}\ &=&\frac 1%
2\tilde{\lambda}(1+\frac J{J_0}).
\end{eqnarray*}
For a given value of $J/J_0$, we have a set of solutions for $\tilde{\lambda}%
,$ $\tilde{\Delta},$ $\tilde{\Delta}_2.$ Our solutions in this paper are
limited to $\min (\tilde{\omega}(k))\neq 0.$ To determine the physical
properties, we evaluate the dynamic correlation functions for the spin $%
S_i^z $, orbital $T_i^z$, and spin-orbital density operators $2S_i^zT_i^z$.
After some algebra, we have 
\begin{eqnarray*}
&&\chi _X(q,\Omega +i\delta )=\frac 18\int \frac{dk}{(2\pi )^d}\left( \frac{%
C_X(k,q)}{\omega (k)\omega (k+q)}-1\right) \\
&&\times \left[ \frac 1{\Omega +i\delta +\omega (k)+\omega (k+q)}-\frac 1{%
\Omega +i\delta -\omega (k)-\omega (k+q)}\right]
\end{eqnarray*}
with ($X=S,T,ST$) 
\begin{eqnarray*}
C_S(k,q) &=&\lambda ^2-a\left( k\right) a(k+q)-b(k)b(k+q); \\
C_T(k,q) &=&\lambda ^2+a\left( k\right) a(k+q)-b(k)b(k+q); \\
C_{ST}(k,q) &=&\lambda ^2-a\left( k\right) a(k+q)+b(k)b(k+q).
\end{eqnarray*}
There are a set of relations 
\begin{eqnarray*}
\chi _S(0,\Omega ) &=&0; \\
\chi _T(0,\Omega ) &=&\chi _{ST}(Q,\Omega ); \\
\chi _T(Q,\Omega ) &=&\chi _{ST}(0,\Omega ); \\
\chi _S(Q,\Omega ) &=&\chi _T(0,\Omega )+\chi _T(Q,\Omega ) \\
&=&\chi _{ST}(0,\Omega )+\chi _{ST}(Q,\Omega ),
\end{eqnarray*}
where for $\Omega >0$ 
\begin{eqnarray*}
\mathop{\rm Im}%
[\chi _T(0,\Omega )] &=&\frac \pi 4\int \frac{dk}{(2\pi )^d}\frac{a^2(k)}{%
\omega ^2(k)}\delta (\Omega -2\omega (k)); \\
\mathop{\rm Im}%
\mathop{\rm Im}[\chi _{T}(Q,\Omega )] &=&\frac \pi 4\int \frac{dk}{(2\pi )^d}%
\frac{b^2(k)}{\omega ^2(k)}\delta (\Omega -2\omega (k)).
\end{eqnarray*}
If the minimum of $\omega (k)$ is non-zero, $%
\mathop{\rm Im}%
[\chi _X(q,\Omega )]$ become non-zero ONLY when $\Omega \geq 2\min (\omega
(k)).$ Thus the collective excitations for the density-density correlation
function have a finite energy gap, $\Delta _{gap}=2\min (\omega (k))$. It is
worth mentioning that the solution has broken the discrete the permutation
symmetry of spin and orbital. This can be seen from the fact that, in
general, $\chi _S(q,\Omega )\neq \chi _T(q,\Omega ).$ The same expressions
are obtained for the second set of solutions if we permute the indices $S$
and $T$. The spectra and free energy as well as the energy gap are identical
to the first set of solutions. Thus the two sets of solutions are
energetically degenerated. The symmetries in the two states are different.
More important, the double degeneracy of the ground state was also observed
in one-dimensions in other approaches.\cite{Kolezhuk98,Pati98} Therefore,
this two-fold degeneracy is not a consequence of the mean field approaches
and can be regarded as an evidence to support our mean field theory.

Now we come to evaluate the energy gap by solving the mean field equations.
On a one-dimensional chain, the energy gap can be evaluated analytically by
introducing a parameter $x_0=(\tilde{\Delta}^2-\tilde{\Delta}_2^2)/(1-\tilde{%
\Delta}_2^2).$ The energy gap and the ratio of $J_0/J$ are 
\begin{eqnarray*}
\Delta _{gap} &=&4z(J+J_0)\times \left\{ 
\begin{array}{c}
\frac{K(x_0)-E(x_0)}{\pi x_0},\text{ if }x_0<0; \\ 
\sqrt{1-x_0}\frac{K(x_0)-E(x_0)}{\pi x_0},\text{ if }x_0\geq 0;
\end{array}
\right. \  \\
\frac J{J_0} &=&-1+2\frac{K(x_0)-E(x_0)}{E(x_0)-(1-x_0)K(x_0)}
\end{eqnarray*}
where $K(x)$ is the complete elliptic integral of the second kind and $E(x)$
is the complete elliptic integral of the first kind. We have established a
one-to-one correspondence between the ratio $J_0/J$ and the energy gap. We
find that there is a turning point at $J_0/J=1$. The theory fails at the
symmetric point $J_0=0.$ The energy gap still opens at that point, which is
in conflict with the solution of Bethe ansatz. \cite{Sutherland75} The same
problem was encountered in the spin-1/2 SU(2) theory in one dimension.

\begin{figure}[tbp]
\centerline{\epsfig{file=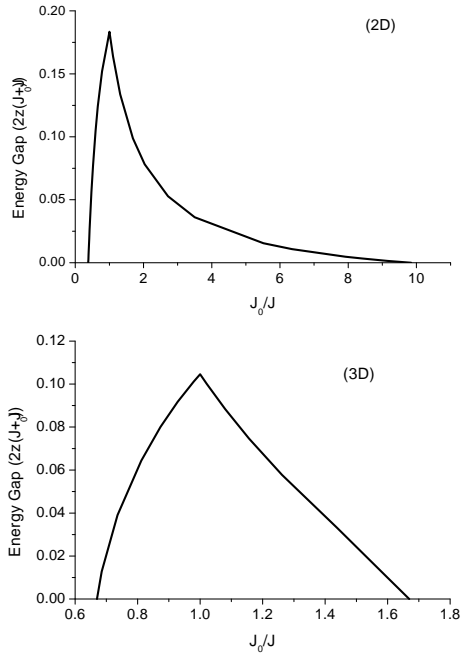, width=7.5cm}}
\caption{The energy gap via the ratio of $J_0/J$ in two- and
three-dimensions.}
\end{figure}

The energy gaps for two- and three-dimensional lattices are plotted in
Fig.1. We find that the energy gap opens in the regime of $0.380<J_0/J<9.84$
for $d=2$ and of $0.666<J_0/J<1.667$ for $d=3$. The gap closes at two
critical ratios $J_0/J.$ From Fig.1, it is shown that the energy gap appears
in a larger parameter range in two-dimension than in three dimension.. This
is consistent with the fact that the quantum fluctuations are stronger in
two dimension. Out of the above parameter regimes the Bose condensations
have to be considered, otherwise the mean field equations have no solutions.
In this case the ground state may possess long-range orders as we discussed
in the symmetric point $J=-J_0$.\cite{Zhang01} We shall discuss these phases
elsewhere. As far as we know the Schwinger boson mean field theory is very
successful for the spin liquid state for $s=1$ in one dimension, and anti-
and ferromagnetic states for higher dimensions. Our theory shows it also
works very well for spin-orbital liquid states in higher dimensions.

The formation of a spin energy gap indicates that the ground state is a
spin-orbital liquid. Experimentally the energy gap can be measured from
magnetic susceptibility. There are several higher-dimensional materials such
as Na$_2$Ti$_2$Sb$_2$O and CaV$_4$O$_9.$\cite{Taniguchi95} It is believed
that the orbital degrees of freedom plays an important role in the formation
of spin gap.\cite{Pati98,Katoh98} The low temperature phases of these
materials may be relevant to the spin-orbital liquid with the energy gap as
we discuss in this paper.

In conclusion we develope a Schwinger boson mean field theory for the
formation of a spin-gap in any dimensions. The ground state of the model is
a spin-orbital liquid with an energy gap in an extensive parameter regime.
This ground state breaks the discrete symmetry of permutation of spin and
orbital, and is doubly degenerated.

This work was supported by a RGC grant of Hong Kong and a CRCG grant of the
University of Hong Kong.

\bigskip

\end{document}